 \documentclass[sigconf]{acmart}
\setcitestyle{numbers,sort&compress}          
\usepackage{multirow}

\usepackage{amsmath}
\usepackage{graphicx}
\usepackage{listings}
\usepackage{hyperref}
\graphicspath{ {./images/} }
\usepackage{float}

\usepackage{float} 
\usepackage{listings}
\usepackage{xcolor}

\usepackage{algorithm}
\usepackage{algorithmic}

\lstdefinestyle{py}{
  language=Python,
  basicstyle=\ttfamily\small,
  keywordstyle=\color{blue!70!black}\bfseries,
  commentstyle=\color{gray!70}\itshape,
  stringstyle=\color{green!40!black},
  showstringspaces=false,
  frame=single,
  framerule=0.4pt,
  numbers=left,
  numberstyle=\tiny\color{gray},
  numbersep=6pt,
  xleftmargin=1em,
  tabsize=2,
  breaklines=true,
  columns=fullflexible
}

\setcopyright{none} 
\acmConference[Anonymous Submission to ? 2026]{Name of acmConference}{2026}{Place of acmConference}
\acmYear{2026}
\begin{document}

\title{Head Count: Privacy-Preserving Face-Based Crowd Monitoring}
\author{Fatemeh Marzani, Thijs van Ede, Geert Heijenk, Maarten van Steen}
\affiliation{%
 \institution{University of Twente}
 \city{Enschede}
  \country{The Netherlands}
}
\email{f.marzani, t.s.vanede, geert.heijenk, m.r.vansteen@utwente.nl}

\begin{abstract}


An important aspect of crowd monitoring is knowing how many people we are dealing with. Sometimes, knowing the size of a crowd in a single location and at a specific moment is enough. Matters become problematic when counting the same people across different locations or counting them over longer periods of time. In those cases, we need to identify and later reidentify a person, which immediately leads to privacy concerns. Until recently, solutions have been based on unique identification of carry-on devices, yet privacy improvements have caused transmitted information to be randomized, rendering this technique mostly useless. We propose to use biometric data instead. We introduce a pipeline that counts people based on face recognition, yet without ever being able to reveal the identity of individuals. 
To count, a camera initially detects a face, extracts its features, and derives an identifier using a fuzzy extractor. The original facial image is then deleted. Identifiers are inserted into homomorphically encrypted Bloom filters. This allows oblivious set membership testing directly on encrypted data, enabling the system to count across locations or across different moments, without revealing any identities. We provide an initial evaluation of our method that shows promising results. 

\end{abstract}


\maketitle

\section{Introduction}

Automated crowd monitoring systems play a crucial role in urban planning, public safety, and event management by analyzing human movements. There are two core tasks in crowd monitoring: footfall and crowd flow. Footfall is the count of detected individuals at a specific location within a defined time window; crowd flow is the more challenging task of counting how many detected people appear across different locations. Footfall can be derived directly from local observations at a specific moment. However, flow estimation requires re-identifying individuals across space and time.\footnote{In this paper, we concentrate on the combination of space \emph{and} time, considering that the situation of re-identification at the same location but at a later moment is just a special case.} This re-identification step introduces significant privacy risks. It requires storing and linking personally identifiable information (PII), which can enable large-scale tracking and unauthorized profiling. Such practices also raise concerns under privacy regulations such as the EU General Data Protection Regulation (GDPR).

Many techniques estimate footfall, from visual density regression~\cite{Counting_without_models} to edge-based systems like Cerberus~\cite{Cerberus}, which runs analytics on-device and exports only counts. However, crowd-flow estimation methods remain limited as they require stable identifiers that persist across locations. For a while, WiFi-based techniques provided such identifiers; mobile devices broadcast probe requests containing fixed MAC addresses, enabling reliable flow estimation. Stanciu et al.~\cite{EBF} proposed a method for privacy-preserving statistical counting using Bloom filters (BFs), probabilistic data structures supporting set operations; together with homomorphic encryption (HE), i.e., a type of encryption that allows performing operations on encrypted data. Their system allows counting over encrypted representations of sets or intersections of sets of devices without revealing what is being counted.\footnote{Intersections are used to determine crowd flow, measuring which individuals are present at \emph{multiple} locations over time.} They have shown that by combining BFs with HE it is possible to provide statistical counts while protecting the data of individuals.
This approach is no longer feasible; modern operating systems now randomly assign MAC addresses, making repeated observations of the same device indistinguishable. Nevertheless, the core insight remains valuable: If we can extract person-unique identifiers from carry-on devices such as smartphones or access cards, accurate and privacy-preserving flow estimation remains feasible.

The core challenge that remains is how to reliably identify and re-identify individuals for counting flows. Instead of relying on carry-on devices, we shift to biometric signals, specifically facial observations captured by cameras. Our focus in this work is on counting how many people move across non-overlapping camera ranges. The main advantage of this shift is that it removes the assumption that individuals carry smartphones, access cards, or other device-based identifiers. Faces are naturally present and can be detected consistently in locations. The drawback is that camera-based systems are heavily dependent on the quality of face detection and by themselves introduce privacy concerns.
This work tackles both problems.
Changes in pose, lighting, occlusion, camera angle, or image quality can make two images of the same person look different.
As a result, the facial embeddings extracted from these images are noisy and unstable: the same person often produces different feature vectors at different locations. This variability makes re-identification much harder than in device-based methods. Combined with privacy issues, it also prevents the direct use of raw facial embeddings as identifiers to count unique individuals and estimate movement between locations.
To overcome these limitations, our approach extracts approximate identifiers (simHashes) from facial images at each location and converts them into stable person-unique identifiers that cannot be traced back to original faces using fuzzy extractors. This places camera-based analytics in the same position as earlier systems that relied on identifiers derived from carry-on devices, but without requiring individuals to have such devices.

Our pipeline consists of several stages. First, face images are processed to extract high-dimensional embeddings that represent the visual appearance of each detected individual. These embeddings are then converted into compact binary strings using SimHash, which preserves the similarity between repeated observations of the same person. Because these binary strings may still vary across captures, we apply a fuzzy extractor to transform them into consistent, non-invertible identifiers. These identifiers can then be inserted into Bloom filters, producing efficient set representations. Finally, Bloom filters are encrypted using fully homomorphic encryption (FHE). The server can then compute the intersection size, telling us how many identifiers appear at two or more different locations directly on the encrypted data.
In this way, our system estimates the size of the crowd flow without revealing who is being counted or exposing facial information. Each stage of this pipeline has its own sources of error, from face detection and embedding extraction to SimHash collisions, fuzzy-extractor reconstruction thresholds, and Bloom-filter false positives. Together, these factors determine the overall accuracy and precision of the final flow estimate. Improving the robustness and reliability of each step is an important direction for further research.

\section{Related Work}
Efforts to preserve privacy by monitoring crowds with face data have focused on footfall counting, which aggregates detections in a single location. Chan et al. ~ \cite{Counting_without_models} use density estimation in visual features to count people without models or tracking, reporting only crowd density (footfall) without addressing flow between locations. Similarly, the Cerberus system ~\cite{Cerberus} processes analytics on-camera, discarding frames, and exporting only aggregate footfall counts rather than crowd flow between locations. Brazauskas et al. ~ \cite{Cerberus} introduce Cerberus, using edge-based face detection to preserve privacy and localization in urban settings, primarily reporting footfall density without flow estimation. Chen et al. ~\cite{chen2025privacy} propose privacy-preserving crowd counting using federated learning, enabling secure model training without sharing face data, but it again targets footfall density rather than flow across locations. Ahmed et al. ~\cite{ahmed2024revolutionizing} explore voice-driven face recognition for crowd surveillance, integrating multimodal data for rapid identification, but their work emphasizes identification over flow estimation. 
Ren et al. ~\cite{ren2024rapoo} develop RAPOO, an efficient privacy-preserving facial expression recognition framework using secure multiparty computation, which is limited to expression analysis and again does not address crowd flow. 

Stanciu et al. \cite{EBF} propose a privacy-preserving crowd-monitoring system using Bloom filters and homomorphic encryption, capable of estimating both crowd density (footfall) and crowd flow across locations. Later, Marzani et al.~\cite{stopWatchingMe} improved privacy in sparsely populated crowd regions within that framework~\cite{EBF} by introducing controlled uncertainty sampling methods (e.g. detection or hash sampling).
However, their approach does not support noisy data, such as face embeddings affected by variations in pose, lighting, or occlusions, limiting its applicability to face-based monitoring.

In contrast, our proposed pipeline extends beyond footfall by estimating crowd flow across multiple locations using face data. By integrating fuzzy extractors with encrypted Bloom filters, we compute aggregate intersection sizes (flow counts) without revealing identities or storing linkable identifiers, addressing a gap left by methods that focus only on density or lack support for noisy biometric data like faces.

\section{Methods}
We consider two cameras, \(A\) and \(B\), deployed at different locations, each detecting faces over time. 
Our goal is to estimate how many people seen at location \(A\) later appear at location \(B\), I.e., our aim is to compute the size of the intersection between the sets of detected faces from both sites—without revealing who those people are.

This task introduces two key challenges.  
First, face data are inherently noisy and repeated captures of the same person differ due to pose, lighting, or camera conditions. To 
reliably compute intersections, we must convert these noisy face embeddings into stable identifiers that remain consistent for the same individual across cameras.  
We address this by using SimHash to preserve similarity between embeddings, and fuzzy extractors to correct small variations and produce deterministic, non-invertible identifiers.

Second, we compute the intersection size without storing, revealing, or comparing any raw identifiers or intermediate data.  
This can be achieved by converting identifiers into compact probabilistic data structures known as Bloom filters and encrypting them using Fully Homomorphic Encryption (FHE). This approach allows set operations to be performed directly on the encrypted Bloom filters.

\subsection{Pipeline Overview}
\begin{figure*}[t]
    \centering
    \includegraphics[width=.8\textwidth]{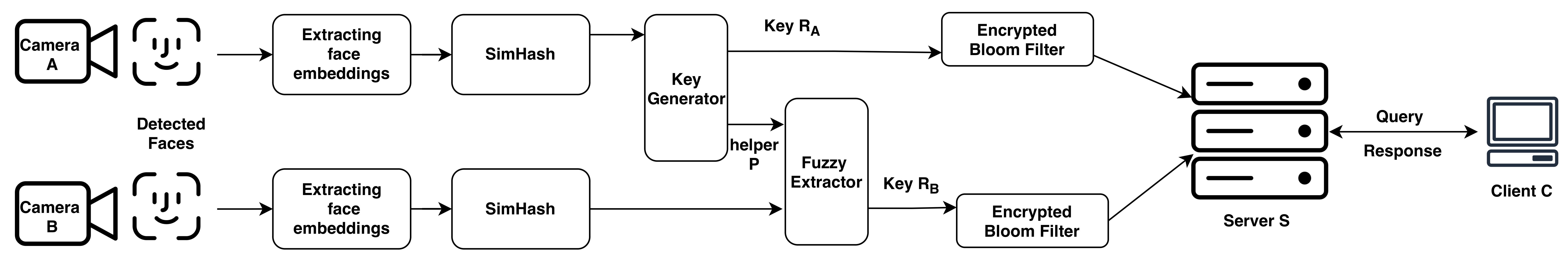}
    \caption{System model of the privacy-preserving face-based crowd monitoring pipeline.}
    \label{fig:system_model}
\end{figure*}

Figure~\ref{fig:system_model} illustrates the setup of the system. The system consists of four entities: two cameras (\(A\) and \(B\)), a central server (\(S\)), and authorized clients (\(C\)).  Each camera detects faces and processes them locally to prevent raw image exposure.  

\textbf{1. Extracting face embeddings.}
We apply a pre-trained deep neural network (DNN) to extract high-dimensional embeddings \(v \in \mathbb{R}^d\) from each detected face.    
This embedding captures the unique visual features of a face, so that similar faces produce similar vectors.
In practice, images of the same person generate embeddings that are close but not identical, while different people produce clearly distinct ones.
To make these embeddings comparable, we first convert them into binary representations that preserve similarity using SimHash.
However, even after this step, small differences may remain between binary strings of the same person.
To address this, we later apply a fuzzy extractor, which can reproduce the same identifier when two binary strings differ only within a limited Hamming distance.

\textbf{2. SimHash for similarity.}
We use SimHash~\cite{simHash}, a locality-sensitive hashing (LSH) method that maps similar inputs to similar binary outputs. SimHash preserves the angular similarity of embeddings by projecting them onto random hyperplanes, ensuring that two captures of the same person produce similar binary codes with small Hamming distance.
Since fuzzy extractors operate in Hamming space, SimHash bridges the gap between continuous face embeddings and discrete binary input.

\textbf{3. Fuzzy extractors for stability and privacy.}
Although SimHash maintains similarity, small visual variations can still flip bits if they are close to a SimHash plane.  
We use fuzzy extractors~\cite{dodis2004fuzzy} to generate a stable, reproducible, and non-invertible identifier from noisy binary inputs. They act as an error-tolerant bridge between biometric variability and cryptographic determinism.
Each extractor produces a key \(R\) and public helper data \(P\).  
If a similar hash reappears on another camera, it reconstructs the same key using \(P\).  
This mechanism ensures consistent identifiers between sites, while preventing any reverse reconstruction of biometric data.

\textbf{4. Bloom filters for compact set representation.}  
We use Bloom filters to compactly represent the individuals detected by each camera while enabling efficient, privacy-preserving computation.  
In our pipeline, the Bloom filter serves as a space-efficient encoding of the stable identifiers produced by the fuzzy extractor. This structure allows each camera to store and later compare large sets of identifiers without ever revealing their actual values. A Bloom filter is a probabilistic data structure that represents a set using a bit array of length \(m\), initially filled with zeros, and \(k\) independent hash functions.  
Each identifier is hashed \(k\) times, and the corresponding bit positions are set to one.  
After inserting all identifiers, the number of bits set to one (\(t\)) provides an accurate estimation of how many unique elements are represented~\cite{Swamidass2007MathematicalCF}.  
The estimated number of inserted elements cardinality, \(c\)) can be computed as:  
\begin{equation}
    c = -\frac{m}{k} \ln \left(1 - \frac{t}{m}\right).
    \label{eq:cardinality}
\end{equation}

Bloom filters also support efficient set operations.  
The intersection between two encoded sets can be approximated with a simple bitwise AND (\&) operation:  
\[
BF_{A \cap B} \approx BF_A \;\&\; BF_B.
\]  
The number of bits set to one in the resulting intersection filter (\(BF_{A \cap B}\)) can then be used in Eq.~(\ref{eq:cardinality}) to estimate \(|A \cap B|\) the number of individuals appearing at both locations.

\textbf{5. Fully Homomorphic Encryption for secure computation.}
Although Bloom filters hide identifiers, their bit patterns could still leak information if transmitted in plaintext as adversaries may test for the presence of known individuals.  
We therefore encrypt each Bloom filter using Fully Homomorphic Encryption (FHE), which allows arithmetic operations on encrypted data.  
The entire system works on an \textbf{epoch} basis, an epoch typically capturing 5 minutes of facial data. In other words, steps 1--5 are repeated after an epoch elapses; the plaintext Bloomfilters are discarded by the respective camera, and only their encrypted versions are stored at the server.

\textbf{6. Decryption and cardinality estimation.}
When the server receives a query from the client, it computes the encrypted intersection of two Bloom filters as:
\[
Enc(t)=\sum_{i=1}^{m}Enc(BF_A[i])\times Enc(BF_B[i]),
\]
where multiplication emulates bitwise AND, and addition counts set bits.  
The server operates only on ciphertexts and learns nothing about the underlying data. The client decrypts the encrypted total \(Enc(t)\) and estimates the number of shared individuals using the Bloom-filter cardinality Formula \ref{eq:cardinality}.

The complete process operates as follows.  
Each authorized client generates a public–private key pair and shares the public key with both cameras.  
During each epoch, cameras \(A\) and \(B\) detect faces and convert them into SimHash binary strings. The 
camera \(A\) applies a fuzzy extractor to generate a stable identifier \(R_A\) along with public helper data \(P_A\). Helper data \(P_A\) are shared with camera \(B\), allowing it to attempt reconstructing the same identifier to match faces observed at its location. The 
camera \(B\) uses these helpers with its own SimHashes to reproduce the corresponding identifiers \(R'_B\) when the Hamming distance between the two hashes falls within the fuzzy threshold.  
Both cameras then encode their final identifiers into Bloom filters, encrypt them using the client’s public key, and send the encrypted Bloom filters to the central server.  
The server performs homomorphic multiplications and additions on ciphertexts to estimate intersections directly in the encrypted domain, without accessing any plaintext data.  
Finally, the client decrypts the aggregate ciphertext, obtains the total number of set bits (\(t\)), and applies Eq.~(\ref{eq:cardinality}) to estimate the number of individuals appearing across both locations.  
This process ensures that only encrypted, aggregated data are processed, while raw facial information and intermediate representations are deleted.

\section{Preliminary Evaluation}
We evaluate the reliability of the fuzzy extractor stage in our privacy-preserving pipeline, which converts noisy SimHash outputs into stable identifiers prior to encrypted Bloom-filter intersection.\footnote{The source code for this work is publicly available at: \url{https://anonymous.4open.science/r/simhash-to-bch-fuzzy-extractor-C12A/README.md}} The objective is to assess whether fuzzy extractors can consistently reproduce identical keys from different images of the same individual. The accuracy of encrypted Bloom-filter intersections has been extensively analyzed in~\cite{EBF} and is therefore not repeated here.

\begin{figure}[t]
    \centering
    \includegraphics[width=\linewidth]{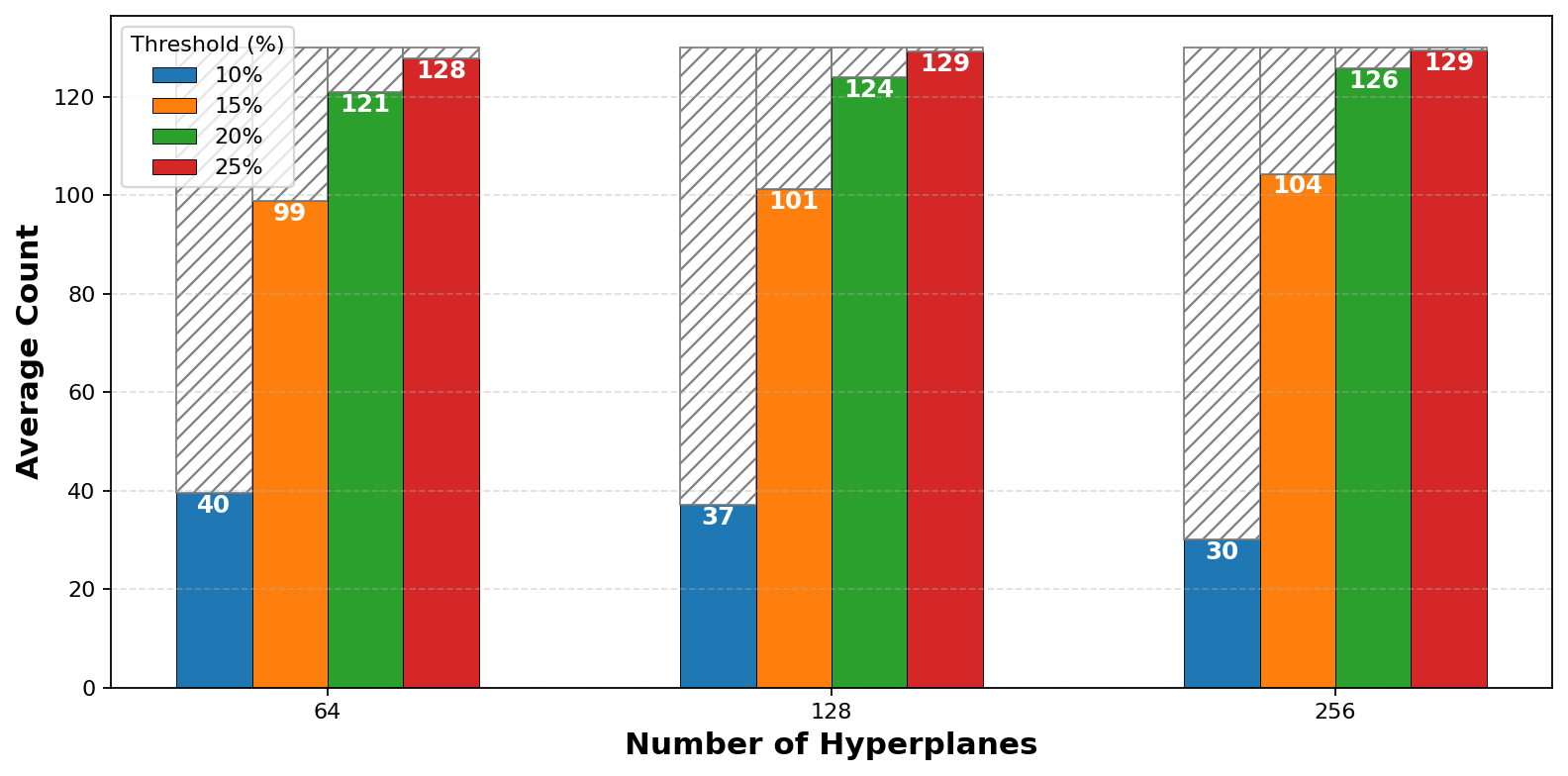}
    \caption{Average TP and FN counts across different SimHash length and fuzzy thresholds.}
    \label{fig:matches}
\end{figure}

We used the SCface dataset~\cite{Grgic2011}, which contains 130 individuals, each captured in nine images under varying lighting conditions and poses, allowing a controlled simulation of cross-camera variability. Faces are detected using VOLO~\cite{volo} and embedded using a ResNet-50 model pre-trained on VGGFace2~\cite{vggface2}. For each identity, four images are randomly assigned to site \(A\) and four to site \(B\), representing realistic sequences of multiple frames of the same individual captured under different conditions. Each camera performs the early stages of the pipeline locally: extracting embeddings, converting them into binary strings via SimHash, and generating stable identifiers using a fuzzy extractor. For each individual, camera \(A\) produces a key \(R_A\) and corresponding public helper data \(P_A\) from the consensus SimHash of its observations. Camera \(B\) then attempts to reconstruct this key using its own SimHash output and the helper data \(P_A\). If reconstruction succeeds for the same identity (\(R'_B = R_A\)), the result is counted as a True Positive (TP); otherwise, as a False Negative (FN). If reconstruction succeeds using helper data of a different identity, it is counted as a False Positive (FP).
We vary the number of SimHash hyperplanes \(n_{\text{bits}} \in \{64,128,256\}\) and the relative error ratio \(r \in \{10\%,15\%,20\%,25\%\}\), where the effective Hamming tolerance is defined as \(\tau = r \cdot n_{\text{bits}}\). we repeated each configuration over 100 random seeds to average out projection noise. Figure~\ref{fig:matches} shows the average TP and FN counts in all configurations. 
With \(n_{\text{bits}} \geq 128\) and relative error ratios \(r \geq 20\%\), nearly all 130 individuals produce consistent keys, demonstrating that the fuzzy extractor effectively corrects the variations of SimHash. Lower error ratios increase FN counts, reflecting reduced tolerance to intra-person variability. These results confirm the reliability of the fuzzy extractor stage in stabilizing similarity-preserving hashes, ensuring that the identifiers passed to the Bloom-filter stage are deterministic and suitable for encrypted intersection. Table~\ref{tab:prf} further quantifies this behavior using precision, recall, and F1-score. At \(r = 10\%\) achieved high precision but low recall, indicating overly strict matching that misses many correct reproductions. Increasing \(r\) significantly improves recall. Longer representations (\(n_{\text{bits}} = 128,256\)) achieve near-perfect precision and recall at moderate error ratios (\(r \geq 20\%\)), showing improved robustness and reduced accidental collisions. 
Overall, these results show that fuzzy extractors reliably stabilize noisy facial embeddings and provide explicit control over the precision–recall trade-off, enabling accurate and privacy-preserving crowd-flow estimation~\cite{EBF}.

\begin{table}[t]
\centering
\caption{Precision, recall, and F1-score for different SimHash lengths and fuzzy thresholds \(r\).}
\label{tab:prf}
\begin{tabular}{c c c c c}
\hline
Hyperplanes & Error Ratio \(r\) & Precision & Recall & F1-score \\
\hline
\multirow{4}{*}{64}
 & 10\% & 1.000 & 0.305 & 0.465 \\
 & 15\% & 0.997 & 0.761 & 0.862 \\
 & 20\% & 0.995 & 0.930 & 0.961 \\
 & 25\% & 0.993 & 0.983 & 0.988 \\
\hline
\multirow{4}{*}{128}
 & 10\% & 1.000 & 0.286 & 0.443 \\
 & 15\% & 1.000 & 0.780 & 0.876 \\
 & 20\% & 1.000 & 0.953 & 0.976 \\
 & 25\% & 1.000 & 0.994 & 0.997 \\
\hline
\multirow{4}{*}{256}
 & 10\% & 1.000 & 0.231 & 0.374 \\
 & 15\% & 1.000 & 0.803 & 0.890 \\
 & 20\% & 1.000 & 0.968 & 0.984 \\
 & 25\% & 1.000 & 0.996 & 0.998 \\
\hline
\end{tabular}
\end{table}

\section{Conclusion and Future Work}
In this paper, we show that it is possible to count how many people appear across two camera locations while preserving privacy and without being able to identify who they are. 
We extend existing privacy-preserving crowd-flow systems that rely on deterministic identifiers to handle noisy biometric data, such as faces. 
Our method generates stable identifiers from facial embeddings using fuzzy extractors, maps them to encrypted Bloom filters, and computes intersection sizes directly on encrypted data. 
Through experiments on the SCface dataset, we demonstrate that fuzzy extractors effectively stabilize noisy face embeddings into deterministic identifiers.

In future work, we will evaluate how sensitive our approach is to different datasets, including images captured under diverse conditions, larger populations, and video-based scenarios. 
We will analyze when and why the method succeeds or fails in better understanding its operational limits and optimize it for real-world, privacy-preserving crowd analytics.

\section{Acknowledgements}
This research is funded by the Dutch Research Council (NWO) through the PERSPECTIEF Program P21-08 'XCARCITY'. We gratefully acknowledge their financial support. 

\citestyle{acmnumeric}
\setcitestyle{numbers,sort&compress}
\bibliographystyle{unsrtnat}

\bibliography{ccs-sample}

\end{document}